\documentclass[fleqn,10pt]{wlscirep}
\usepackage{amsmath}
\usepackage{amsfonts}
\usepackage{amssymb}
\usepackage{verbatim}
\usepackage{graphicx}
\usepackage{color}

\title{Strength of weak layers in cascading failures on multiplex networks: case of the international trade network}
\author[1]{Kyu-Min Lee}
\author[1,*]{K.-I. Goh}
\affil[1]{Department of Physics and Institute of Basic Science, Korea University, Seoul 02841, Korea}
\affil[*]{corresponding author: kgoh@korea.ac.kr}

\begin{abstract}
 Many real-world complex systems across natural, social, and economical domains consist of manifold layers to form multiplex networks. The multiple network layers give rise to nonlinear effect for the emergent dynamics of systems. Especially, weak layers that can potentially play significant role in amplifying the vulnerability of multiplex networks might be shadowed in the aggregated single-layer network framework which indiscriminately accumulates all layers. Here we present a simple model of cascading failure on multiplex networks of weight-heterogeneous layers. By simulating the model on the multiplex network of international trades, we found that the multiplex model produces more catastrophic cascading failures which are the result of emergent collective effect of coupling layers, rather than the simple sum thereof. Therefore risks can be systematically underestimated in single-layer network analyses because the impact of weak layers can be overlooked. We anticipate that our simple theoretical study can contribute to further investigation and design of optimal risk-averse real-world complex systems.
\end{abstract}

\begin{document}
\flushbottom
\maketitle

\section*{Introduction}
\label{intro}
No systems in nature and society are in complete isolation. Rather, they are constantly interacting with others through a myriad of different interactions. For instance, in human society, individuals interact one another through several social relationships such as friendship, kinship, co-workership, {\it etc.} Such ``multiplexity'' in complex systems~\cite{NoN,multilayer,multilayer2,epjb} induces nontrivial emergent behaviors such as layer-wise structural features \cite{szell,kmlee2012,bianconi}, catastrophic cascades of  failure~\cite{buldyrev,sonsw,viability,multiple_transitions}, facilitated social cascades~\cite{brummitt,osman,lee2014threshold} and modulated information spreading~\cite{diffusion,min}, to name but a few.

Global economy can also be best viewed as the multiplex system, in which each country is connected to other countries through various politico-financial channels such as commodity trading, capital trading, political alliance, and the like~\cite{nations}. 
The degree of multiplexity in global economy has become more important than ever, owing to the recent fast development of information and communication technology which enabled unprecedentedly more active and sophisticated interaction channels between different economic sectors, making the whole system increasingly more interwoven and interdependent. 
This increasing connectivity and interdependence within the global economic system has elevated the possibility of systemic risks that could not be accounted for by orthodox economic theory, as many have argued recently~\cite{buffett, unconference, haldane_may, battiston, kaushik}. 
Subprime mortgage crisis in 2008 and the european sovereign-debt crisis in late 2009 would be prime recent manifestation of such risks. In order to  better estimate, understand, and predict the risk of such large-scale economic turmoils, therefore, new perspective on crisis modeling based on multiplex networks beyond the single-network framework is called for \cite{schweitzer, battiston2012, controllability,catanzaro, battiston2013, caldarelli, galbiati, thurner,distress-propagation,garas2008,saracco}.  

\begin{figure}[t]
\includegraphics[width=14cm]{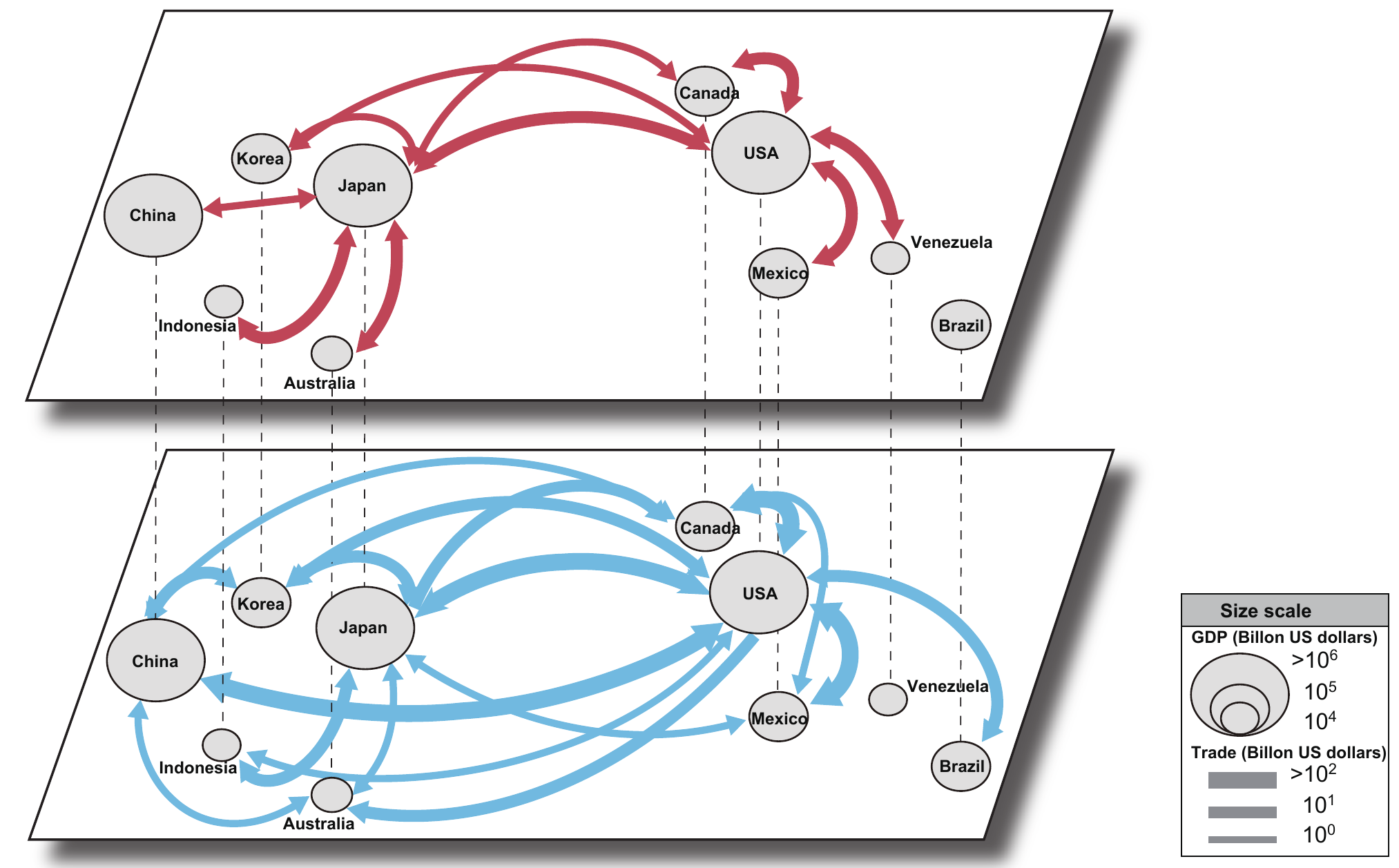}
\caption{Schematic example of the multiplex global trade network of $10$ countries. The first layer indicates the primary industry sector (red arrows) and the second layer represents the secondary industry sector (blue arrows).
Only trading relations with above $5$ Billion US dollars are selected as links for the visibility. When there are reciprocal relations, the volume of the links is obtained by summing up the volume of two directed links. Each node's size is scaled as its GDP volume.}
\label{fig:cartoon}
\end{figure}

Cascading failures on networks is one of the most actively studied classes of problems in network dynamics~\cite{sandpile,MotterLai,dynamic-book,watts2002}, which can serve as the theoretical platform for crisis modeling \cite{kmlee2011}. In this paper, we study a simplistic crisis spreading model on the multiplex network of international trade. 
In so doing, our objective is twofold. Our first and foremost objective is to better understand the role of multiplexity in cascading failure dynamics on multiplex networks in general. Secondly, through the study of multiplex crisis spreading model we shall hope to draw some insight in the financial context at the theoretical level. 
To this end, we construct the multiplex network of two layers of international trade, based on the empirical trading relations among countries in the primary and the secondary industry sector, respectively (Fig.~\ref{fig:cartoon}), on which the multiplex crisis-spreading model is simulated (see Model and Methods sections for details). The rationale underlying this construction is that these two layers constitute an interdependent relationship in the sense that the proper operation of the primary industry sector is bound to the stable demand from the secondary industry sector, the proper operation of which in turn relies on the stable supply of raw materials from the primary sector. A crucial consequence of such interdependency in multiplex system is that the failure of a node in one layer also causes the failure of that node in the other layer, which could dramatically amplify the entangled cascade of failure \cite{buldyrev,brummitt}. We note that the complete functional interdependency between layers in our model is an obvious oversimplification of reality. Primary aim of such simplification is to elucidate the basic role of the interdependency and the effect of weak layer most clearly. More detailed and sophisticated modeling of  interdependency relation by extensions of our model should be desirable as future works especially to address more detailed and specific economic contexts.

From theoretical perspective of multiplex network dynamics, another key property of multiplex trade network 
is that the two layers have significantly different density and weight of connections in the two layers:
The secondary industry sector layer is similarly dense ($1.06$ times mean degree in terms of the number of trading partner countries) yet carries $\approx3.8$ times larger weight (in terms of the monetary volume of trades) than the primary industry sector.
Therefore, when one were to aggregate the two layers in the hope of simplifying the dynamics, one tends to end up with wrong prediction by neglecting the cooperative multiplex coupling between the two layers to which the weak layer can exert just as substantial an effect.
To address this issue more concretely, we assess individual country's potential impact on the avalanche dynamics with a simplified cascading failure model on the multiplex trade network and argue for the strength of weak layer by comparing the multiplex dynamic outcome with the corresponding single-layer counterpart. 
We find that the multiplex coupling facilitate the cascades of failure to a great extent compared with equivalent dynamics on the aggregated single-layer network. The emergent multiplex dynamics is further shown to be more than sum of individual layer dynamics. It will finally be shown that the primary industry layer can play as instrumental role as the secondary industry layer in driving the multiplexity-induced cascades, despite its significantly smaller weight. All these indicate that the risk of cascades could be considerably underestimated in the aggregated single-layer framework, even when the link weight is largely dominated by a single layer, thus shadowing the potential elevated risk of catastrophic failure. 

\section*{Multiplex cascading failure model}
\label{subsec:model}
In this section we introduce the multiplex cascading failure model on the weighted and directed network with heterogeneous nodal capacity. The present model is a multiplex generalization of the single-layer cascading failure model studied in ref.~\citen{kmlee2011}.
In the multiplex cascading failure model, each node engages in two different layers of links and the two layers are considered interdependent each other. Due to the interdependence, a node's failure in one layer implies its failure in the other layer. A brief description of the dynamic rule of the model is as follows (for a full algorithmic presentation, see Methods section). When a node $i$ fails, it decreases the weights of its all outgoing links weights by the fraction $f_{1}$  and $f_{2}$ in layer $1$ and $2$ respectively, such that
\begin{align}
W_{ij}^{(1)} \gets (1-f_{1})W_{ij}^{(1)} \quad\text{and}\quad W_{ij}^{(2)} \gets (1-f_{2})W_{ij}^{(2)}, \quad j \in \{\textrm{neighbors of node }i\}.
\label{link_decrement-multiplex}
\end{align} 
The same weight reduction is applied to all its incoming links, $W_{ji}^{(1)}$ and $W_{ji}^{(2)}$, as well. 
The weight reduction parameters $f_1$ and $f_2$ are independent free parameters of the model with the range of $f_1, f_2 \in [0, 1]$. 
The link weight reduction induced by the node failure affects the neighboring nodes so that  
an unfailed neighboring node $k$ would also fail if the following failure condition is met: It fails if the total decrement of link weights to and from it exceeds in any layers the fraction $t_1$ in layer $1$ or $t_2$ in layer $2$ of its nodal capacitance $C_{k}$, that is,
\begin{eqnarray}
\textrm{max}(\sum_{l\in F} f_1W^{(1)}_{kl}, \sum_{l\in F} f_1W^{(1)}_{lk}) > t_{1}C_{k} \quad \textrm{or} \quad \textrm{max}(\sum_{l \in F} f_2W^{(2)}_{kl}, \sum_{l\in F} f_2W^{(2)}_{lk}) > t_{2}C_{k}.
\label{eq:multiplex_rule}
\end{eqnarray} 
where $F$ is the set of failed nodes. If the failure condition is met, the node $k$ fails and the cascade of failure proceeds and the dynamics proceeds in discrete steps by following the parallel update. The cascade of failure initiated by small fraction of seed nodes is terminated at the step at which there are no more newly-failed nodes. The total number of failed nodes in the cascade process is called the avalanche size $A$, which is the main observable for this model.
The failure tolerance parameters $t_1$ and $t_2$ are also independent free parameters of the model with the range of $t_1, t_2 \in [0,1]$. But the two parameter sets $\{f_1,f_2\}$ and $\{t_1,t_2\}$ are not independent; the dynamics depends only on their fractional combination\cite{kmlee2011} , $\phi_1=f_1/t_1$ and $\phi_2=f_2/t_2$, thus leaving only two independent free parameters $\{\phi_1,\phi_2\}$ in the model, and the failure condition becomes 
\begin{eqnarray}
\textrm{max}(\phi_1\sum_{l\in F} W^{(1)}_{kl}, \phi_1\sum_{l\in F} W^{(1)}_{lk}, \phi_2\sum_{l \in F} W^{(2)}_{kl},  \phi_2\sum_{l\in F} W^{(2)}_{lk}) > C_{k}.
\label{eq:multiplex_rule}
\end{eqnarray} 

To assess the effect of multiplexity, we compare the outcomes of multiplex cascade model with the corresponding single-network counterpart, which is constructed by aggregating the two layers so that the link weight $W^{(\textrm{s})}_{ij}$ is given by 
\begin{align}
W^{(\textrm{s})}_{ij} = W^{(1)}_{ij} + W^{(2)}_{ij}, 
\label{eq:simplex-sum}
\end{align}
where the superscript $(\textrm{s})$ denotes `simplex-network.'
Since the weight of links in simplex network is obtained by the summing over two layer's weights, the effective control parameter $\phi_{\text{eff}}$ is given by
\begin{align}
\phi_{\text{eff}} = \frac{f_1 w_1 + f_2 w_2}{t_1 + t_2}
\label{eq:effective}
\end{align}
where $w_1$ and $w_2$ are the fraction of total weights, 
$w_1 = W_{\textrm{total}}^{(1)} / (W_{\textrm{total}}^{(1)}+W_{\textrm{total}}^{(2)})$ and $w_2 = W_{\textrm{total}}^{(2)} / (W_{\textrm{total}}^{(1)}+W_{\textrm{total}}^{(2)})$. Therefore, the condition of failure of a node $k$ in the simplex network is given by
\begin{align}
\textrm{max}(\phi_{\text{eff}}\sum_{l\in F}W_{kl}^{(s)}, \phi_{\text{eff}}\sum_{l\in F}W_{lk}^{(s)}) > C_k
\end{align}
where $F$ is the set of failed nodes. Then we compare the outcome with the multiplex network results. 

\section*{Multiplex international trade network}
The international trade network is one of the representative examples of weighted and directed networks, which also has multirelational property \cite{barigozzi, mastrandrea}. Thus it is an ideal testbed on which the multiplex cascading failure model can be simulated to investigate the effects of multiplexity and the disparity in layer strengths. 
The international trade data~\cite{comtrade} allows the link categorization in terms of commodity classes up to nearly a hundred of `trade layers,' as described in ref.~\citen{barigozzi, mastrandrea}. Here we take an alternative,  intermediate or mesoscopic level classification of primary and secondary industry sectors, under the rationale of layer interdependency as posited in the Introduction. 
In this setting of multiplex international trade network, the nodes denote individual countries and the link weights $W^{(a)}_{ij} (W^{(a)}_{ji})$ denote the trade volume that a country $i$ exports (imports) to (from) a country $j$ in the $a$-ary industry sector layer. 
An important property of this international trade network is that the total weight (trade volume) of primary industry layer ($\approx$1.3 Trillion US dollars) is much smaller than that of secondary industry layer ($\approx$4.9 Trillion US dollars). Consequently, if one were to aggregate the two layers into one, the role of primary industry layer would likely be obscured and underestimated. 

For the nodal capacity $C_i$ to tolerate the impact of link weight change induced by neighboring countries' failure, a natural quantity to use would be the Gross Domestic Product (GDP) being a proxy for the country's economic potential capacity~\cite{imf}. More detailed description of the trade and GDP data are provided in Methods section.

\begin{figure}[t]
\includegraphics[width=.8\textwidth]{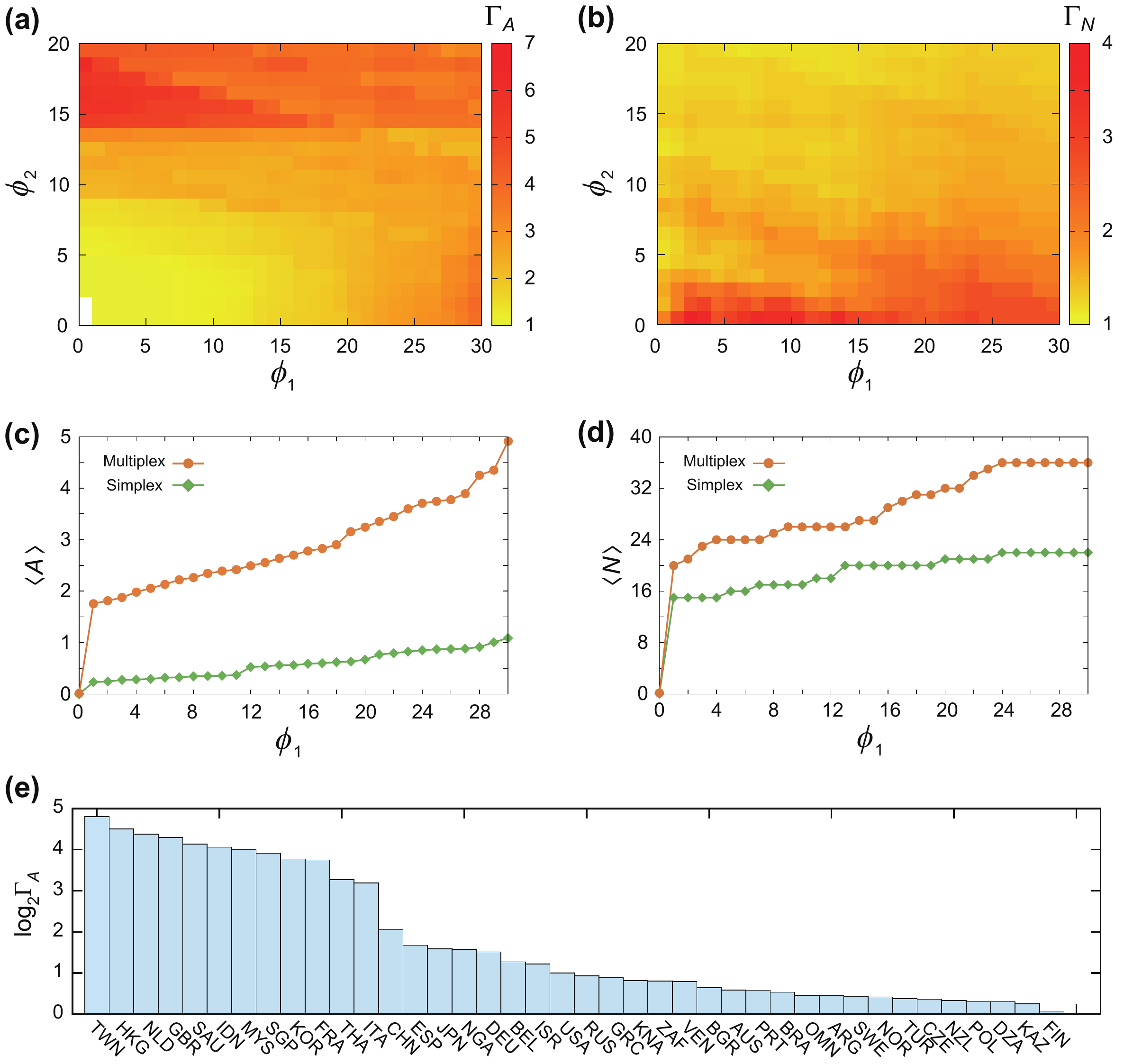} 
\caption{The ratio between the average avalanche size of multiplex ($\langle A_{m}\rangle$) and single-network model ($\langle A_{s}\rangle$) (a), and the average number of countries with nonzero avalanche size of multiplex ($N_m$) and single-network model ($N_s$) (b) with varying control parameters, $\phi_1$ and $\phi_2$. The white area indicates the value of $\Gamma$ is exactly $1$, so the average avalanche size of multiplex and simplex are strictly equal. For the fixed parameter $\phi_2 = 10$, the average avalanche size (c) and the number of nonzero avalanche size countries (d) according to the different $\phi_1 $ for multiplex (orange circles) and single-network (green diamonds) model are presented. (e) Rank-ordered bar-plot of the avalanche size ratio ($\Gamma_A$) of countries with $\Gamma_A > 1$.} 
\label{fig:amvsas-general}
\end{figure}

\section*{Results}
\label{sec:results}

\subsection*{Multiplex coupling facilitates the cascading failure dynamics} 
\label{subsec:facilitation}

The main observable of the cascading failure model is the avalanche size $A_i$~\cite{kmlee2011} which is defined as the total number of failed nodes starting from the initial failure of the single country $i$. It can be considered as a measure of potential impact of the country $i$ on the cascading failure dynamics. We compute the avalanche sizes $\{A_i\}$  for all the countries for the given range of control parameters $\phi_1$ and $\phi_2$ for the primary and secondary layer, respectively. As economic crises accompany a few percentage of GDP falls, we choose the reasonable parameter range of $0\le \phi_1 \le 30$ and $0\le \phi_2 \le 20$. For the control parameter $\phi < 1$ in each layer, a country fails only when the volume of trading relation with its failed neighbor is greater than its GDP. Such case is very rare in our empirical data and even so it is found that no further cascades propagate from such cases. Hence we mainly focus our analysis on the case $\phi \ge 1$. In 2009 eurozone crisis, for example, there were significant GDP decrement of some countries from $5$ to $15\%$\cite{bbc}.
From it, we derive the average avalanche size over all epicentric countries $\langle A\rangle = \frac{1}{{\cal N}}\sum_i A_{i}$, where ${\cal N}=160$ is the total number of countries in the data, for the multiplex model (denoted as ${\langle A\rangle}_{\text{m}}$) and for the aggregated single-network model using the effective parameter $\phi_{\text{eff}}$ (denoted as ${\langle A\rangle}_{\text{s}}$). We also compute the number of epicentric countries that can induce additional failure of other countries (that is, $A_i>1$), to be denoted as $N_{\text{m}}$ for the multiplex and $N_{\text{s}}$ for the single-network case.  
These quantities can be interpreted as the likelihood ($N$) and the expected extent ($\langle A\rangle$) of cascading failures from a random epicenter.

By using these two quantities we address the amplifying impact of multiplex coupling in the cascading failure dynamics. 
Specifically, we quantify it by taking the ratios, 
\begin{align} 
\Gamma_A={\langle A\rangle}_{\text{m}}/{\langle A\rangle}_{\text{s}} \quad\text{and}\quad  \Gamma_N=N_{\text{m}}/N_{\text{s}}, 
\end{align} 
respectively.
As shown in Figs.~2a,b, these ratios are obtained to be consistently greater than unity, indicating the generic amplifying effect of multiplex coupling. 
Not only the ratios but we also observe the systematic persistence of the absolute gap in $A_{\text{m}}$ and $A_{\text{s}}$ and that for $N$ in the investigated range of $\phi$ (Fig. 2c,d). 
For the given parameter region, we have found that no countries has negative logarithmic value of $\Gamma$'s, such that the average value of the avalanche size of multiplex model are never smaller than the measures of single-layer counterpart. In the figure 2e, we presented forty countries which have $\Gamma_A>1$ with rank-ordered.

Top-rankers in the avalanche size ratio are overrepresented with industrialized countries in east asian region such as Taiwan, Hong Kong, Indonesia, and Malaysia. The significant amplification of the avalanche size ratio of those countries may stem from their close trading dependencies with Singapore which is well known trading hub in world economy, which were not fully accounted for by the single-layer analysis. 
G8 countries (England, France, Italy, Japan, Germany, USA, Russia, except Canada) make the appearance in the list, as well as China. Holland and Saudi Arabia make the third and fifth place, reflecting their role in entrep\^ot trading and oil trading, respectively.

\begin{figure}[t]
\includegraphics[width=.85\textwidth]{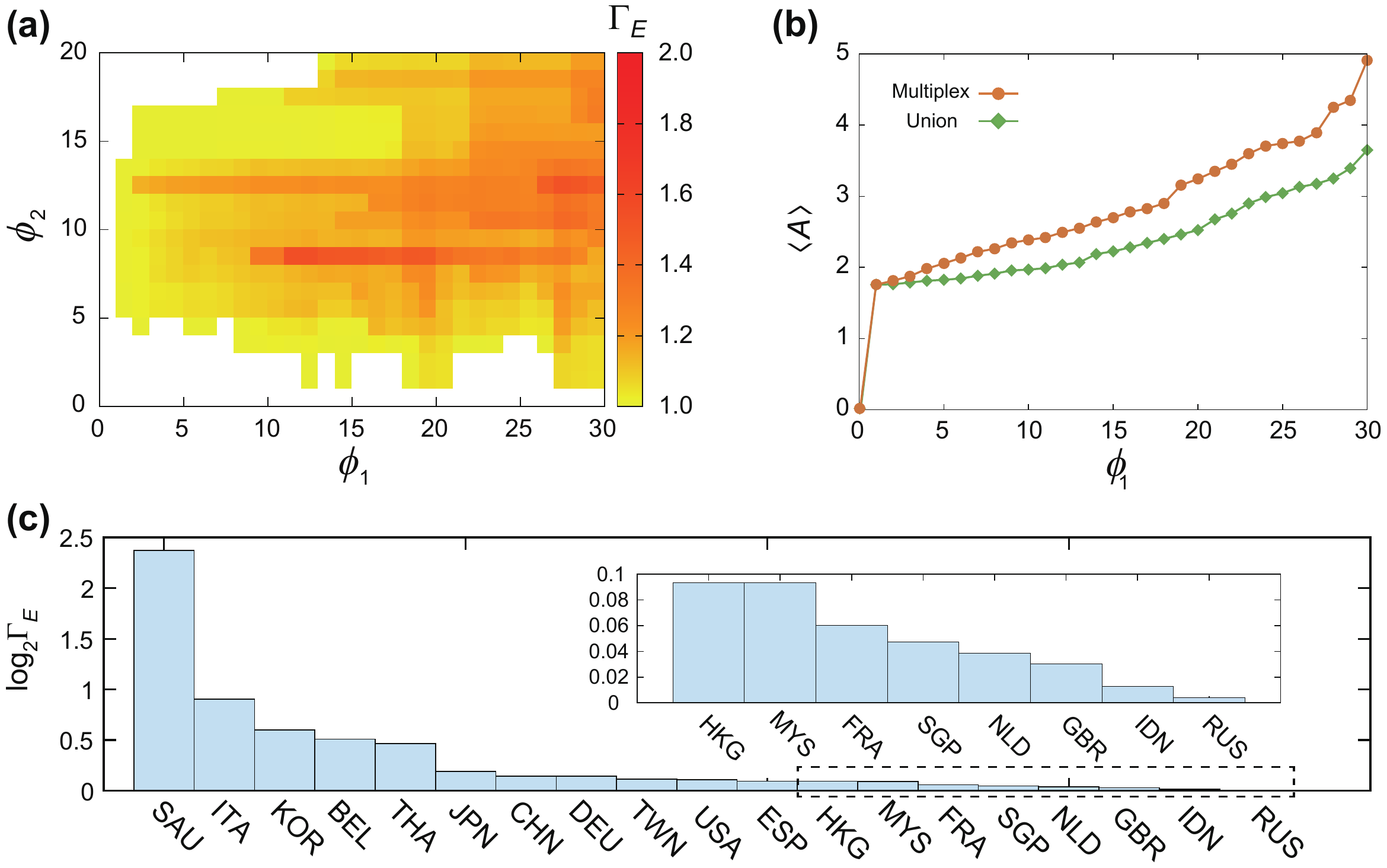} 
\caption{(a) The excess avalanche ratio ($\Gamma_E$) with varying control parameters, $\phi_1$ and $\phi_2$. The white area indicates the value of $\Gamma_E$ is exactly $1$, so the average avalanche size of multiplex and union are strictly equal. (b) For the fixed parameter $\phi_2 = 10$, the average excess avalanche size according to the different $\phi_1 $ for multiplex (orange circles) and single-network (green diamonds) model are presented. (c) The histogram of excess avalanche ratio ($\Gamma_E$) of countries with $\Gamma_E > 1$.} 
\label{fig:adist}
\end{figure}

\subsection*{Multiplex dynamics is more than sum of individual-layer dynamics} 
We verify that the facilitation of avalanche dynamics is cooperative effect of multiplexity rather than simple additive effect of dynamics in isolation. To see this, suppose a hypothetical situation that a country $i$ induces failures of its neighbors $j$ and $k$ in layer $1$, and $l$ and $m$ in layer $2$ respectively. Then the simple union of collapsed countries from a country $i$ is $\bigcup_{\text{collapsed}, i} = \{j, k, l, m\}$ with size $4$. We call this as the union avalanche size of the country $i$, $A_{\text{u}, i}$, and this measures the simple additive effects of each isolated dynamics. 
Similarly to the previous part, we take the ratio of the average avalanche size of multiplex model $\langle A\rangle _{\text{m}}$ and the average union avalanche size $\langle A\rangle_{\text{u}}$, to be called the excess avalanche ratio, 
\begin{align}
\Gamma_{E} = \langle A\rangle_{\text{m}} / \langle A\rangle_{\text{u}},
\end{align}
which gives the measure of more-than-additive, cooperative effect from multiplexity. 

As shown in Fig.~3, the cooperative effect is prevalent. The excess ratio exceeds unity in wide parameter range (Fig.~3a) and the gap between $\langle A\rangle_{\text{m}}$ and $\langle A\rangle_{\text{u}}$ broadens with $\phi$ (Fig.~3b). 
Similar amplification of global cascades from cooperative effect between multiplex layers was also obtained in threshold cascade dynamics~\cite{brummitt,osman,lee2014threshold}. 
Saudi Arabia stands as the significant top-rank country for excess avalanche ratio, followed by Italy, South Korea, Belgium, Thailand, and so forth (Fig.~3c). Expectedly, all the nineteen countries with $\Gamma_E>1$ are also found in the list of countries with $\Gamma_A>1$ in Fig.~2e. The overwhelming excess avalanche ratio of Saudi Arabia may reflect its major role as the oil supplier. Although the oil industry can put biggest and widest influences on most of the secondary industry compared with other natural resources and crude materials in primary sector, the single-layer analysis does not account for this effect properly. Given that we have used the trade data from the year 2000, it would be interesting to further study how its significant role in cascading dynamics would have become by the recent oil price drop and the oversupply in oil market with the production of shale gas.

\begin{figure}[t]
\includegraphics[width=15.5cm]{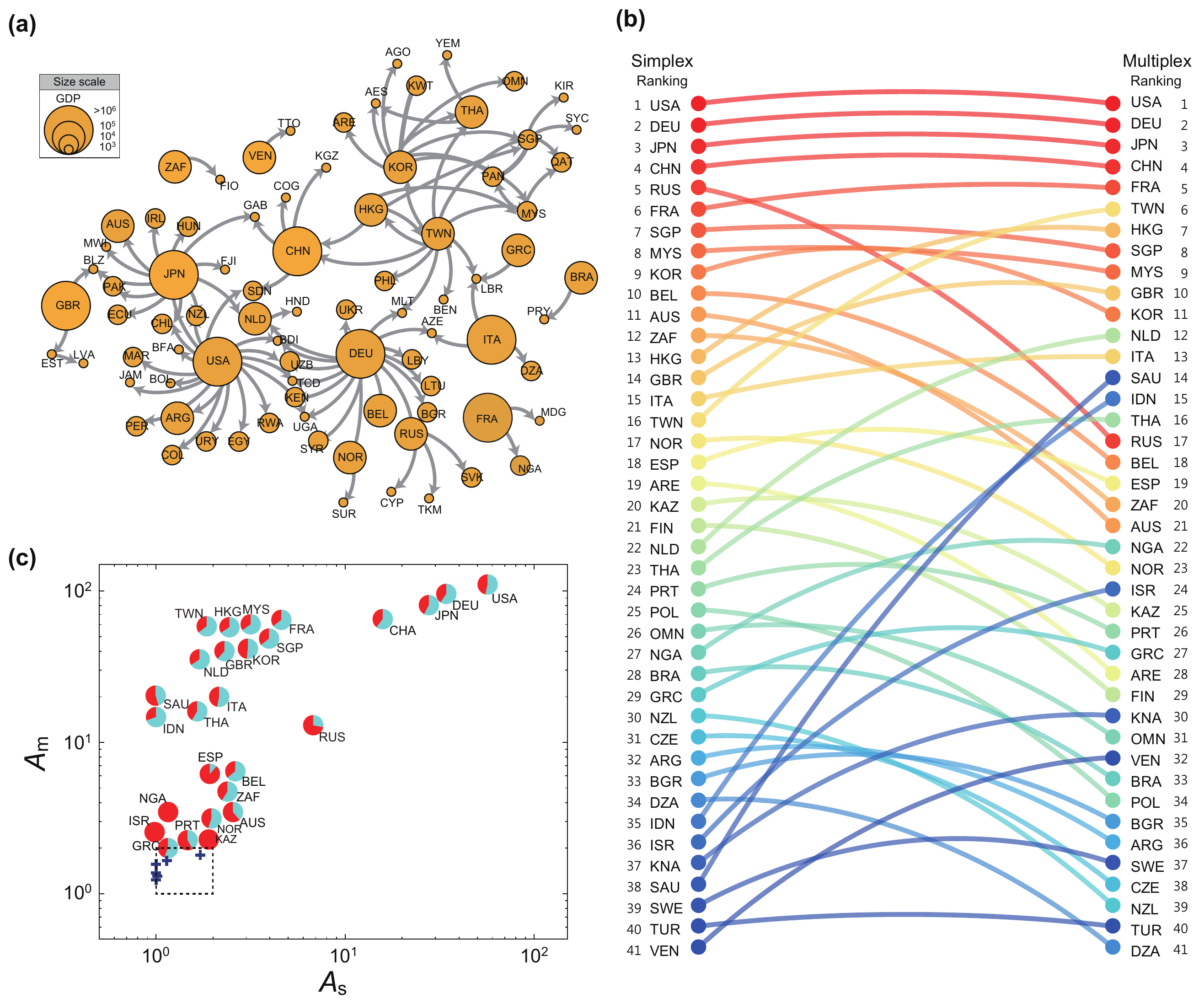} 
\caption{(a) The diagram of avalanche relations ($A\to B$ means $A$ makes collapse $B$) which appear in multiplex model but not in single-network one. Top $100$ frequently happened relations are only presented. (b) Comparison between the ranking of countries based on the avalanche size in single-network and multiplex model. (c) Scatter plot of the avalanche size of single-layer ($A_s$) and multiplex ($A_m$) for individual countries. The proportions of cascades due to the primary (red) and secondary (blue) industry layers are shown as piecharts. Countries with small avalanche sizes of $1 \le A_m \le 2$ and $1 \le A_s \le 2$ (inside of the dashed box) are just presented as blue cross symbols.}
\label{fig:adiff-histogram}
\end{figure}

\subsection*{Strength of weak layers: Role of the primary industry sector}
The multiplexity-induced cascade relations (that is, the failure in multiplex dynamics of a country that did not fail in aggregated single-layer dynamics) constitute complex web structure (Fig.~4a). Even though the figure only presents top one hundred frequently appeared relations, interesting loop (reciprocal) relations, such as the one between Taiwan and Hong Kong, can be found. The existence of the loop in this excess avalanche relation means strongly interdependent mutual economic relationship between those countries, which is not properly captured in the single-layer analysis. Furthermore, the net effect of multiplex coupling is not a simple overall amplification of cascades. Rather, its effect is heterogeneous across the countries, and thus can lead to the active rearrangement of the avalanche size ranking of countries (Fig.~4b). Notably, Saudi Arabia shows biggest leap likely due to its important role as the oil supplier, while Russia, Australia, and South Africa undergo biggest drops in their ranking resulting from their relatively smaller amplification compared with other countries that show big leaps.

What can account for such an amplification effect of the multiplex dynamics compared to the single-layer dynamics? Given the disproportionate difference in link weights (trading volume), might most effect be attributed to the secondary industry layer? The key observation of this study suggests that it is not necessarily so. In fact, significant part of the difference in $A_{\text{m}}$ and $A_{\text{s}}$ must be attributed to the role of primary industry layer. To see this, we show in Fig.~4c the piecharts showing the proportions of cascades due to the primary and secondary industry layers for each country, scatterplotted in ($A_{\text{s}}$, $A_{\text{m}}$) plane. It shows a good share of cascades proceed through the primary industry layer, much larger than expected from the link weights (which is about 20\%).
This shows that the role of the primary industry layer, despite being weak, is not negligible, which would nevertheless be overlooked when the two layers were to get aggregated. 

\begin{figure}[t]
\includegraphics[width=16.6cm]{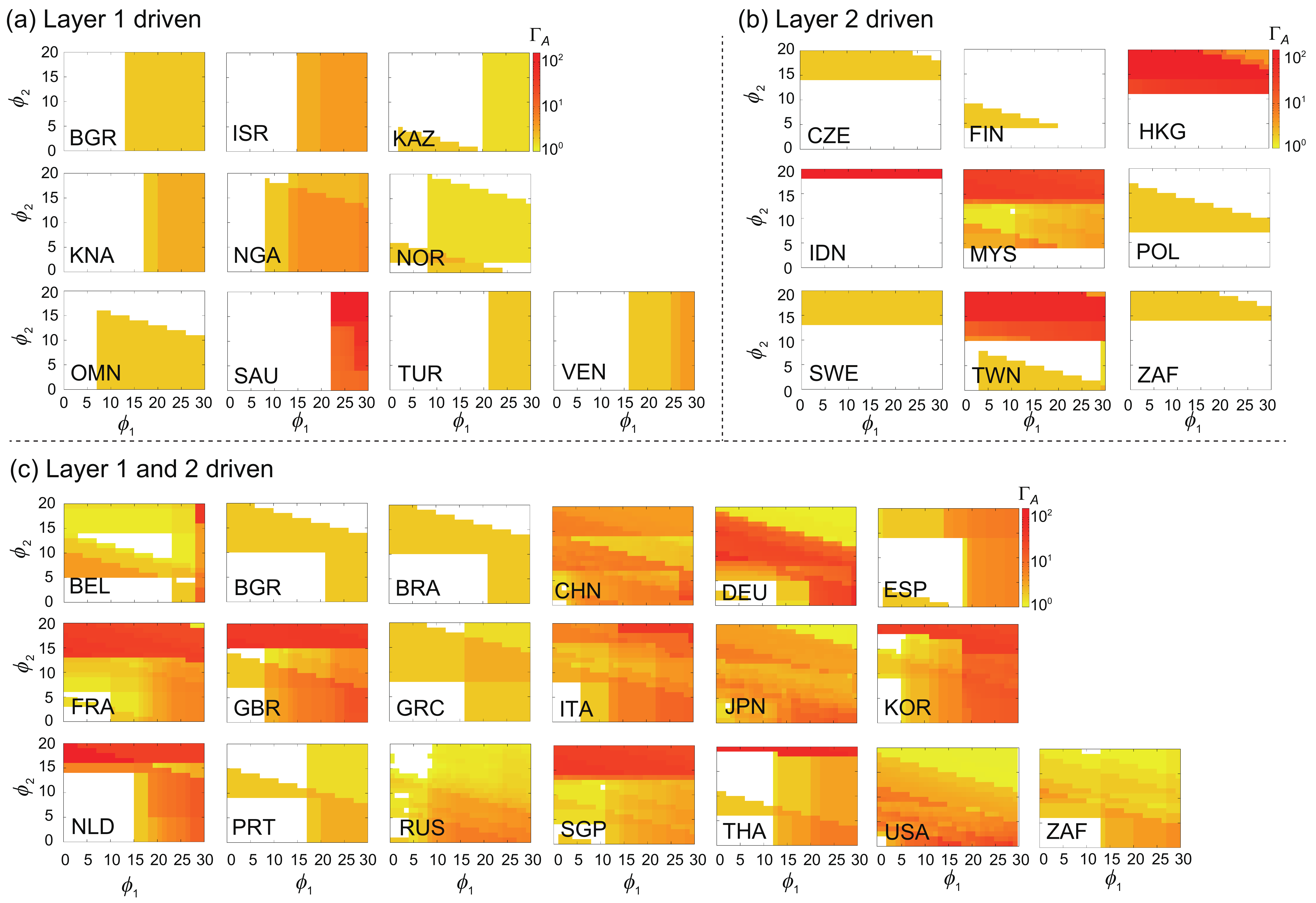} 
\caption{The ratio between the average avalanche size of multiplex and single-network model ($\Gamma_A$) for all countries with $\Gamma > 1$. All countries can be classified as three types according to which layer drives the avalanche dynamics: Layer $1$ driven (a), Layer $2$ driven (b), and both layers driven (c).}
\label{fig:tgp}
\end{figure}

To further quantify the role of each layer in emergent multiplex dynamics, we classified all the countries into three classes by examining their $\Gamma_A$ profiles (Fig.~5). The first class is called layer-1 driven and displays $\Gamma_A>1$ if $\phi_1$ exceeds the threshold, without depending much on $\phi_2$ (Fig.~5a). Ten countries are assigned to this class and typical examples are Saudi Arabia and Israel. On the contrary, the second class displays $\Gamma_A>1$ if $\phi_2$ exceeds the threshold without much depending on $\phi_1$ (Fig.~5b), so it called layer-2 driven. Nine countries are assigned to this and Indonesia and Hong Kong are representative examples. The rest of countries comprise the third class displaying mixed behavior, depending both on $\phi_1$ and $\phi_2$ (Fig. 5c). This classification is not intended to be rigorous but instructive to show that the two layers, despite their significant difference in weights, play comparable role in driving emergent multiplex dynamics.

\section*{Discussion}
\label{sec:discussion}
In this paper we have investigated the problem of cascading failures on multiplex networks, with particular emphasis on the role of weak layers with small weights. A simple theoretical model is simulated on the duplex international trade network with two {\it functionally interdependent} layers representing different industry sector: the primary and secondary industry layers.
The multiplex coupling is found to greatly facilitate the cascades of failure, leading to larger and more frequent avalanche compared with that for equivalent dynamics on the aggregated single-layer network. The emergent multiplex dynamics is also more than sum of individual layer dynamics. In driving such multiplexity-induced cascades the primary industry layer can play as instrumental role as the secondary industry layer, despite its smaller weight by the factor almost four. All these indicate that the risk of cascades could be severely and systematically underestimated when we consider the system in the aggregated single-layer framework, even when the link weight is dominated by a particular layer, thus shadowing the risk of catastrophic failure. 

Real-world complex systems, including the global economic and financial system, is becoming ever more complex. New layers of interaction (new financial instruments or new social interaction medium, say) are invented continuously and get entangled with the existing systems, which itself  are interacting and interdependent one another. Such increasing complexity and interdependency can potentially amplify the risk as is exemplified in this work. And this can be done even when the new layers start very weakly, as long as the system bears collateral nature in it. 
Although the model studied in this work is highly simplified and applied only to one example system of international trade network, the message is expected to be generic. Features specific to the international trade network include the finite size (about two hundreds nodes), network correlations \cite{interlayer,reis,nicosia,jykim,overlap-bianconi,overlap-csf} (manifest as economic bloc and geopolitical clustering), and  fluctuations (year-to-year fluctuation of trades and GDPs) \cite{riad-imf}. Another limitation of the current study is that here we have assumed, quite deliberately, that the layers are completely functionally interdependent and thus collateral-damaged. It is an obvious oversimplification of reality, and can be remedied by more detailed and sophisticated modeling of dependence relationships like partial and/or directional interdependency as well as the more complicated dynamic rules of the multiplex model \cite{brummitt,osman,lee2014threshold,distress-propagation}. We anticipate that our study would help build better understanding and optimal strategy for robustness of the global economy system. A related aspect of international trade network in this regard is its strong positive interlayer degree correlation (with the Pearson correlation between degrees of the country in the two layers being as large as $0.98$), a feature which was shown to increase the robustness of multiplex systems under random failure \cite{reis,interlayer, parshani2010}. 
More detailed studies need be followed to fully address this question in a more proper realistic context, along the line of recent efforts towards designing safer and optimal complex systems. \cite{brummitt-overload, schneider, morone}.

\section*{Methods}
\label{sec:methods}
\subsection*{Algorithm of multiplex cascading failure model}
\begin{enumerate}
\item[1:] Start T=0
\item[2:] A country $i$ collapses
\item[3:] Reduce the weights of all links of in both layers of country $i$ by a fraction $f_{1}$ and $f_{2}$ respectively
\item[4:] \textbf{while} the number of newly collapsed countries $>0$ \textbf{do}
\item[5:] $\quad$ $T$ $\gets$ $T+1$
\item[6:] $\quad$ Reduce the weights of all links of in both layers of countries collapsed at time T by a fraction $f_{1}$ and $f_{2}$
\item[7:] $\quad$ \textbf{for} every non collapsed countries \textbf{do}
\item[8:] $\quad$  $\quad$ \textbf{if} total decrement of either export or import link weights exceeds a fraction $t$ of its GDP in Layer $1$ or $2$?
\item[9:] $\quad$ $\quad$  $\quad$  \textbf{then} the country $j$ collapses
\item[10:] $\quad$  $\quad$\textbf{end if}
\item[11:] $\quad$\textbf{end for}
\item[12:] \textbf{end while}
\item[13:] Record: Avalanche size $A_{i}$
\end{enumerate}
Following the collapse of each country, the avalanche size $A$ is obtained as the number of all collapses of the dynamics.

\subsection*{Data}
The multiplex world trade is based on the real trading relations data between countries and Gross Domestic Product (GDP) data. The GDP is the representative indicator of each countries' economic development. We use them as each countries' capacity of bearing from shocks and economic crises. The GDP data was obtained from the International Monetary Fund (IMF) World Economic Outlook Databases (WEO)~\cite{imf}. From this database we use the GDP value of each countries in the year of $2000$.

The relations between countries are given by the commodity-specific trading data, which is obtained from the UN COMTRADE database~\cite{comtrade}. This dataset provides trading relations among each countries for both exports and imports with specific commodities in every year from $1962$. It provides several commodity classification and we used the SITC Rev.$2$\cite{comtrade} among them. Here we also use the data of the year of $2000$.

Using this classification, we construct the multiplex global trade network with two layers. The first layer (layer $1$) consists of the primary industry products with classification numbers $0$ through $4$. The secondary industry produces with classification numbers from $5$ through $8$ compose the second layer (layer $2$). The classification number $9$ commodities are not classified elsewhere in the SITC Rev.$2$ and can be ignored since their proportion to the total trade volume is negligible. Only countries which are included in both the GDP and trade data are considered so that there are $160$ countries in the multiplex global trade network. For comparing with the effect of multiplex network, we also constructed the single-layer version of global trade network by aggregating two layers.

\section*{Acknowledgments}
The authors would like to thank Prof. Jeho Lee for helpful comments. This work was supported by the National Research Foundation of Korea (NRF) grants funded by the Korea government (MSIP) (No.\ 2011-0014191 and No.\ 2015R1A2A1A15052501).

\section*{Author Contributions}
K.-M.L.\ and K.-I.G.\ conceived the study. K.-M.L.\ executed the research and performed the detailed analysis. K.-M.L.\ and K.-I.G.\ wrote the manuscript. 

\section*{Competing financial interests}
 The authors declare no competing financial interests.

\end{document}